\documentclass[aps,pra,amssymb,twocolumn,superscriptaddress,10pt,tightenlines]{revtex4-1}

\usepackage{graphicx}
\usepackage{dcolumn}
\usepackage{bm}
\usepackage{color}
\usepackage{subfigure}
\usepackage{url}
\usepackage{hyperref}
\usepackage{epstopdf}
\usepackage{soul}

\begin{document}

\newcommand{\beq}{\begin{equation}}
\newcommand{\eeq}{\end{equation}}
\newcommand{\barr}{\begin{eqnarray}}
\newcommand{\earr}{\end{eqnarray}}

\def\bra#1{\langle{#1}|}
\def\ket#1{|{#1}\rangle}
\def\sinc{\mathop{\text{sinc}}\nolimits}
\def\cV{\mathcal{V}}
\def\cH{\mathcal{H}}
\def\cT{\mathcal{T}}
\def\e{\mathrm{e}}
\def\i{\mathrm{i}}
\renewcommand{\Re}{\mathop{\text{Re}}\nolimits}
\newcommand{\tr}{\mathop{\text{Tr}}\nolimits}

\title{Diffraction-limited plenoptic imaging with correlated light}

\author{Francesco V. Pepe}\email{francesco.pepe@ba.infn.it}
\affiliation{INFN, Sezione di Bari, I-70126 Bari, Italy}

\author{Francesco Di Lena}
\affiliation{Dipartimento Interateneo di Fisica, Universit\`a degli studi di Bari, I-70126 Bari, Italy}
\affiliation{INFN, Sezione di Bari, I-70126 Bari, Italy}

\author{Aldo Mazzilli}
\affiliation{Dipartimento Interateneo di Fisica, Universit\`a degli studi di Bari, I-70126 Bari, Italy}

\author{Eitan Edrei}
\affiliation{Fischell Department of Bioengineering, University of Maryland, College Park MD 20742 USA} 

\author{Augusto Garuccio}
\affiliation{Dipartimento Interateneo di Fisica, Universit\`a degli studi di Bari, I-70126 Bari, Italy} 
\affiliation{INFN, Sezione di Bari, I-70126 Bari, Italy}
\affiliation{Istituto Nazionale di Ottica (INO-CNR), I-50125 Firenze, Italy}

\author{Giuliano Scarcelli}
\affiliation{Fischell Department of Bioengineering, University of Maryland, College Park MD 20742 USA} 

\author{Milena D'Angelo}\email{milena.dangelo@uniba.it}
\affiliation{Dipartimento Interateneo di Fisica, Universit\`a degli studi di Bari, I-70126 Bari, Italy} 
\affiliation{INFN, Sezione di Bari, I-70126 Bari, Italy}
\affiliation{Istituto Nazionale di Ottica (INO-CNR), I-50125 Firenze, Italy}

\begin{abstract}
\noindent Traditional optical imaging faces an unavoidable trade-off between resolution and depth of field (DOF). To increase resolution, high numerical apertures (NA) are needed, but the associated large angular uncertainty results in a limited range of depths that can be put in sharp focus. Plenoptic imaging was introduced a few years ago to remedy this trade off. To this aim, plenoptic imaging reconstructs the path of light rays from the lens to the sensor. However, the improvement offered by standard plenoptic imaging is practical and not fundamental: the increased DOF leads to a proportional reduction of the resolution well above the diffraction limit imposed by the lens NA.  In this paper, we demonstrate that correlation measurements enable pushing plenoptic imaging to its fundamental limits of both resolution and DOF. Namely, we demonstrate to maintain the imaging resolution at the diffraction limit while increasing the depth of field by a factor of $7$. Our results represent the theoretical and experimental basis for the effective development of the promising applications of plenoptic imaging.
\end{abstract}

\maketitle

\noindent 
Plenoptic imaging (PI) is a novel optical method for recording visual information \cite{adelson}. Its peculiarity is the ability to record both position and propagation direction of light in a single exposure. PI is currently employed in the most diverse applications, from stereoscopy \cite{adelson,muenzel,levoy}, to microscopy \cite{microscopy1,microscopy2,microscopy3,microscopy4}, particle image velocimetry \cite{piv}, particle tracking and sizing \cite{tracking}, wavefront sensing \cite{thesis_wu,eye,atmosphere1,atmosphere2} and photography, where it currently enables digital cameras with refocusing capabilities \cite{website,ng}. The capability of PI to simultaneously acquire multiple-perspective 2D images brings it among the fastest and most promising methods for 3D imaging with the available technologies \cite{3dimaging}. Indeed, high-speed and large-scale 3D functional imaging of neuronal activity has been demonstrated \cite{microscopy4}. Furthermore, first studies for surgical robotics \cite{surgery}, endoscopic application \cite{endoscopy} and blood-flow visualization \cite{piv2} have been performed. 

The key component of standard plenoptic cameras is a microlens array inserted in the native image plane, that reproduces repeated images of the main camera lens on the sensor behind it \cite{adelson, ng}. This enables reconstruction of light paths, employed, in post-processing, for refocusing different planes, changing point of view and extending depth of field (DOF) within the acquired image. However, a fundamental trade-off between spatial and angular resolution is naturally built in standard plenoptic imaging. If $N_{\mathrm{tot}}$ is the total number of pixels per line on the sensor, $N_x$ the number of microlenses per line, and $N_u$ the number of pixels per line associated with each microlens, then $N_x N_u = N_{\mathrm{tot}}$. Essentially, standard PI gives the same resolution and DOF one would obtain with a $N_u$ times smaller NA. The final advantage is thus practical rather than fundamental, and is limited to higher luminosity (hence SNR) of the final image and parallel acquisition of multi-perspective images.

Correlation plenoptic imaging (CPI) has recently been proposed for overcoming this fundamental limit \cite{cpi_prl}. The main idea is to exploit the second-order spatio-temporal correlation properties of light to perform spatial and directional detection on two distinct sensors: Using correlated beams \cite{cpi_prl,cpi_qmqm,cpi_technologies}, high-resolution ``ghost" imaging is performed on one sensor \cite{pittman,gatti,laserphys,valencia,scarcelliPRL} while simultaneously obtaining the angular information on the second sensor. As a result, the relation between the spatial ($N_x$) and the angular ($N_u$) pixels per line, at fixed $N_{\mathrm{tot}}$, becomes linear: $N_x + N_u = N_{tot}$ \cite{cpi_prl}.

In this paper, we present the first experimental realization of CPI. Our CPI scheme has higher DOF and higher resolution than traditional PI; compared to conventional imaging, it maintains the diffraction-limited resolution but has a $7$ times larger DOF. Therefore, CPI truly pushes imaging to the fundamental limits imposed by the wave nature of light. Our proof-of-principle experiment indicates that CPI can enhance the potentials of PI, paving the way towards its promising applications, especially in situations where the fast acquisition typical of PI needs to be accompanied by high resolution, such as microscopy and 3D imaging. In fact, compared to other 3D imaging techniques, CPI has the advantage of not requiring neither scanning methods (as in confocal microscopy), nor delicate interferometric techniques (as in holography and ptychography), or fast pulsed illumination (as in time-of-flight imaging) \cite{tracking,bookmicroscopy,kim2010principles,hansard2012tof}. 

A schematic representation of the experimental setup is reported in Figure \ref{fig:setup}; technical details are in Appendix C. Based on the ghost imaging phenomenon \cite{valencia,scarcelliPRL}, intensity correlation measurement between each pixel of $S_a$ and the whole sensor $S_b$, described by the Glauber correlation function \cite{scully}, enables retrieving an image of the object on the plane of $S_a$. Such ``ghost'' image is focused provided the distance $z_a$ between the source and the sensor $S_a$ is equal to the distance $z_b$ between the source and the object \cite{valencia,scarcelliPRL}: Due to the spatio-temporal correlation properties of chaotic light, the light source plays the role of a focusing element, and replaces the lens of a standard imaging system characterized by an image magnification $m=1$ \cite{scarcelliPRL}. This justifies the name of {\it spatial sensor} for detector $S_a$, despite it detects a light beam that has never passed through the object. Like standard imaging, both the maximum achievable resolution set by the diffraction-limit ($\Delta x^f$) and the DOF of the ghost image are expected to be defined by the numerical aperture of the focusing element (here, the chaotic light source), as seen from the object ($\mathrm{NA}$). In our case: $\Delta x^f=\lambda/\mathrm{NA}= 14 \:\mu\mathrm{m}$ and, for objects at the resolution limit, $\mathrm{DOF} =\lambda/\mathrm{NA}^2=0.37\,\mathrm{mm}$. In our experiment, the pixel size is chosen to be comparable with the maximum achievable resolution: $\delta x=7.2\:\mu \mathrm{m} \approx \Delta x^f/2$, thus enabling imaging at the diffraction limit. 

To understand how CPI enables increasing the DOF of the acquired image and changing the viewpoint (as required for 3D imaging), let us study the role of the high-resolution sensor $S_b$. Each pixel of this sensor corresponds to the source point from which the detected signal has been emitted. Correlation measurements between pixels of $S_a$ and $S_b$ may thus enable tracing ``light rays'' by joining each object point with each source point \cite{cpi_prl, cpi_qmqm}. Therefore, the high resolution of $S_b$ does not inhibit retrieval of the (ghost) image of the object on $S_a$, but simply provides displaced coherent images, one for each source point (see Appendix A). The conventional (incoherent) ghost image can be recovered by summing the correlations over the whole sensor $S_b$, which corresponds to using the typical ``bucket'' detector of ghost imaging.

The refocusing capability of CPI is governed by the resolution of the source image retrieved by $S_b$, which is defined, together with the numerical aperture $\mathrm{NA}_{b}$ of the lens $L_b$, by the diffraction of light at the object (see Appendix A) \cite{cpi_prl,cpi_qmqm}. In our experiment, both the resolution limit defined by the lens ($\lambda/\mathrm{NA}_{b} = 14 \:\mu\mathrm{m}$), and the pixel size of $S_b$ ($\delta u = 72 \:\mu\mathrm{m}$), have been chosen in such a way that the resolution on the source plane $\Delta u$ is mostly defined by diffraction at the object ($\lambda z_b/a$, where $a$ is the length scale of the smallest details of the object). We thus operate in a regime where imaging performances are limited by the wave nature of light, and not by the microlens size as in standard PI.

\begin{figure}
\centering
{\includegraphics[width=0.45\textwidth]{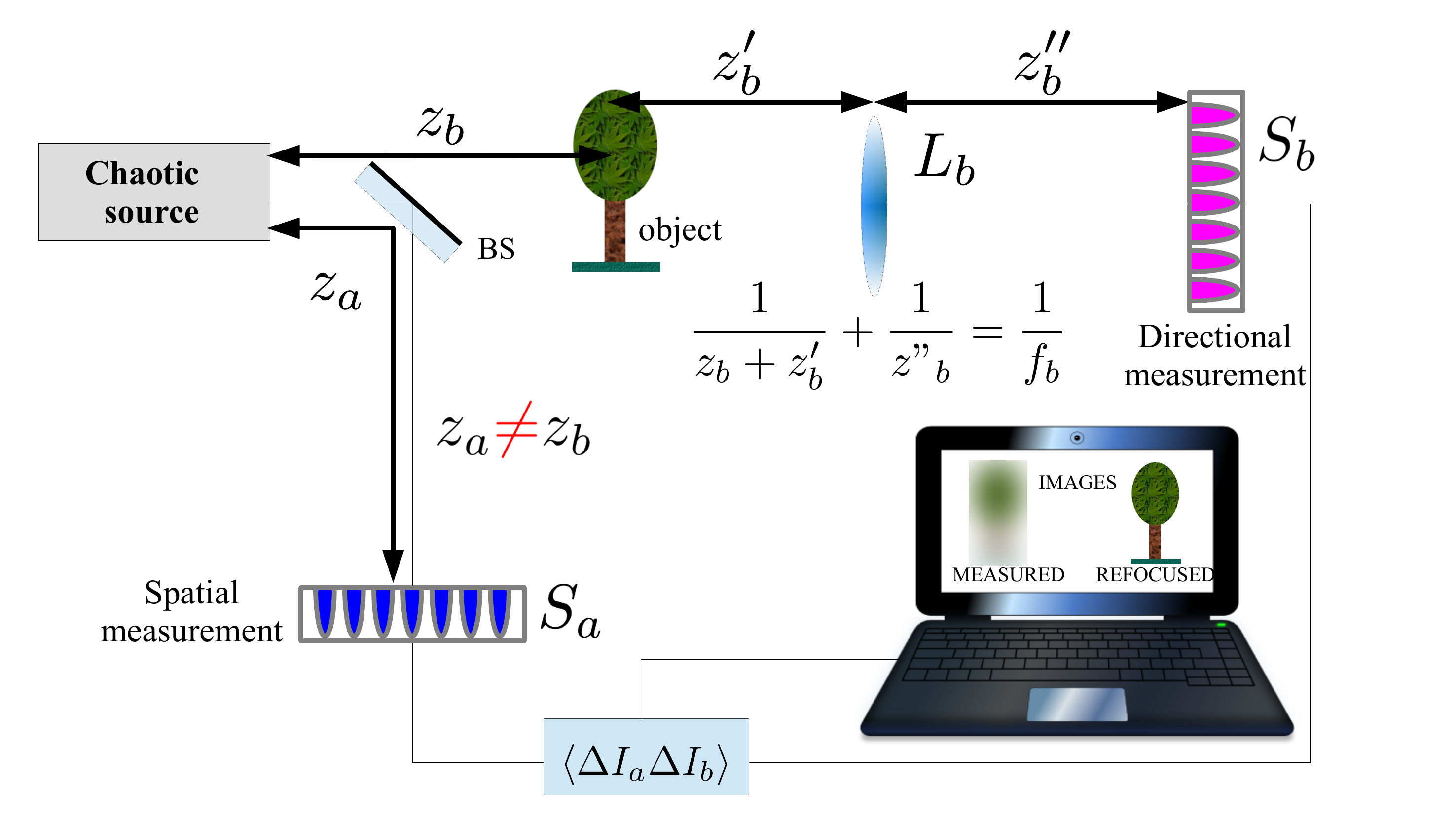}}
\caption{Schematic representation of the experimental setup employed for correlation plenoptic imaging. The lens $L_b$ replaces the whole microlens array of standard PI.}\label{fig:setup}
\end{figure}

In our experiment, we have employed a test target to mimic small details and easily monitor the image resolution, both in the out-of-focus and in the refocused image. In Figure \ref{fig:exp}, we report the experimental results obtained for element 3 of group 2: the three slits have center-to-center distance $d = 0.198 ~\mathrm{mm}$ and slit width $a=d/2$. In the left column, we report the out-of-focus image obtained on $S_a$ by measuring correlation with the whole detector $S_b$, when the mask is placed significantly out of the focused plane ($z_b-z_a \gtrsim 20 \mathrm{mm}$); this is equivalent to the blurred image any conventional imaging system, characterized by the same NA as our CPI scheme, would retrieve at the given defocusing distance. In the right column, we report the same image after implementing the CPI refocusing algorithm \cite{cpi_prl}
\begin{equation}\label{refocus}
\Sigma^{\mathrm{ref}}_{z_a,z_b} (\bm{\rho}_a)\! := \!\int \!
d^2\!\bm{\rho}_b \Gamma_{z_a,z_b} \!\left(\frac{z_a}{z_b} \bm{\rho}_a \! -
\frac{\bm{\rho}_b}{M} \!\left( 1- \frac{z_a}{z_b} \right)\!,
\bm{\rho}_b \right),
\end{equation} 
where $\Gamma_{z_a,z_b} (\bm{\rho}_a,\bm{\rho}_b)$ represents the measured correlation of intensity fluctuations $\langle \Delta I_a \Delta I_b \rangle$ between point $\bm{\rho}_a$ on $S_a$ and point $\bm{\rho}_b$ on $S_b$. The refocusing capability of CPI clearly appears from Figure \ref{fig:exp}, based on  the enhanced resolution and contrast of the refocused image. 

\begin{figure}
\centering 
\includegraphics[width=0.45\textwidth]{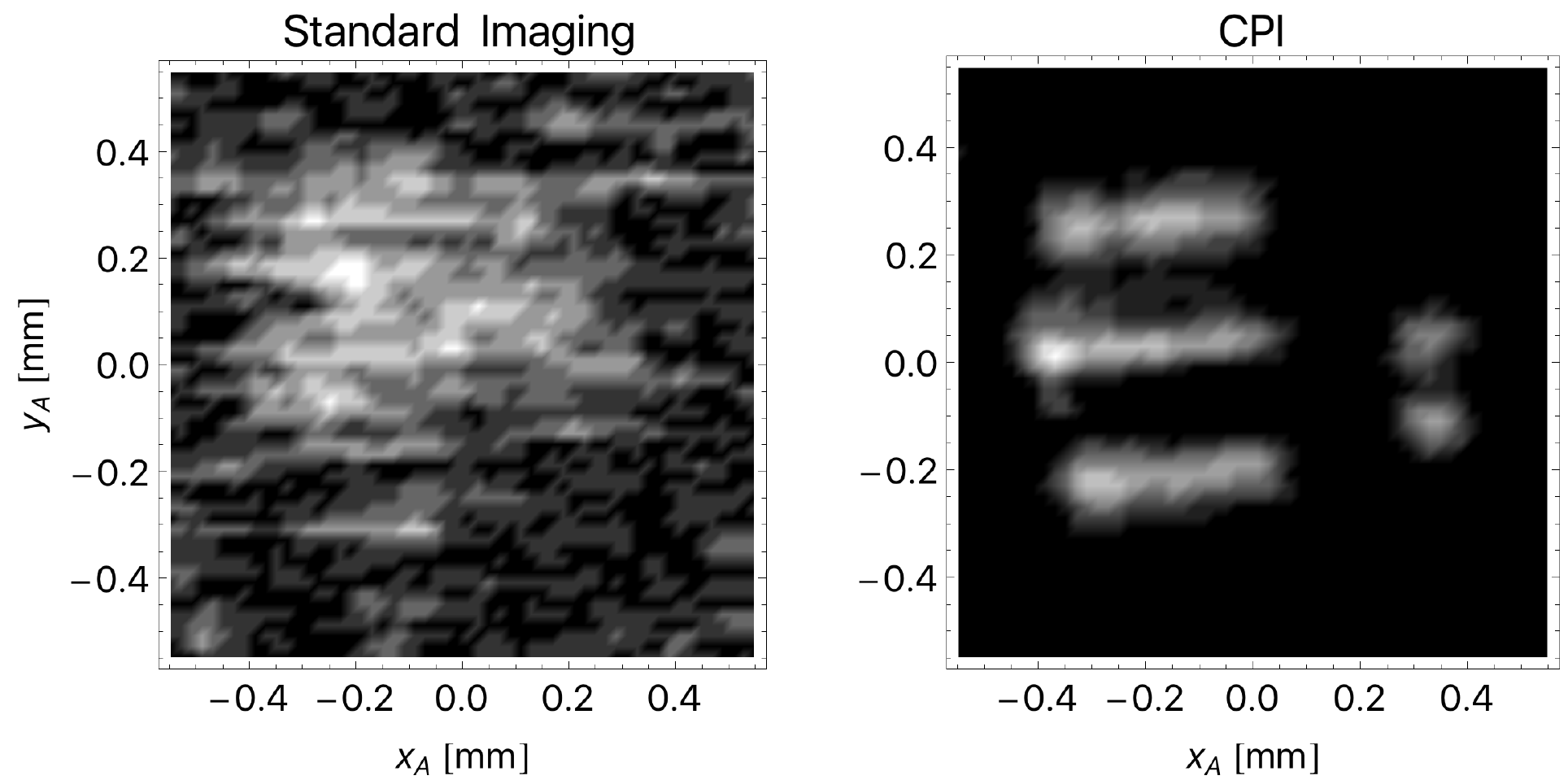}
\caption{Comparison between the experimental out-of-focus image obtained by placing the object (element 3 of group 2 of a test target) in $z_b - z_a = 21 \,\mathrm{mm}$ (left column), and the corresponding CPI refocused images (right column). This experimental scenario is denoted as measurement B in Figure \ref{fig:visib}. The experimental data are taken with a pixel size at the diffraction limit ($\delta x = 7.2\:\mu\mathrm{m}$), while the refocused image has a pixel size scaled by a factor $z_b/z_a$, in line with Eq.~(\ref{refocus}). After correlation measurement, low-pass Gaussian filtering and thresholding in the Fourier domain was applied to remove uncorrelated background. }\label{fig:exp}
\end{figure}

\begin{figure}
\centering
\includegraphics[width=0.45\textwidth]{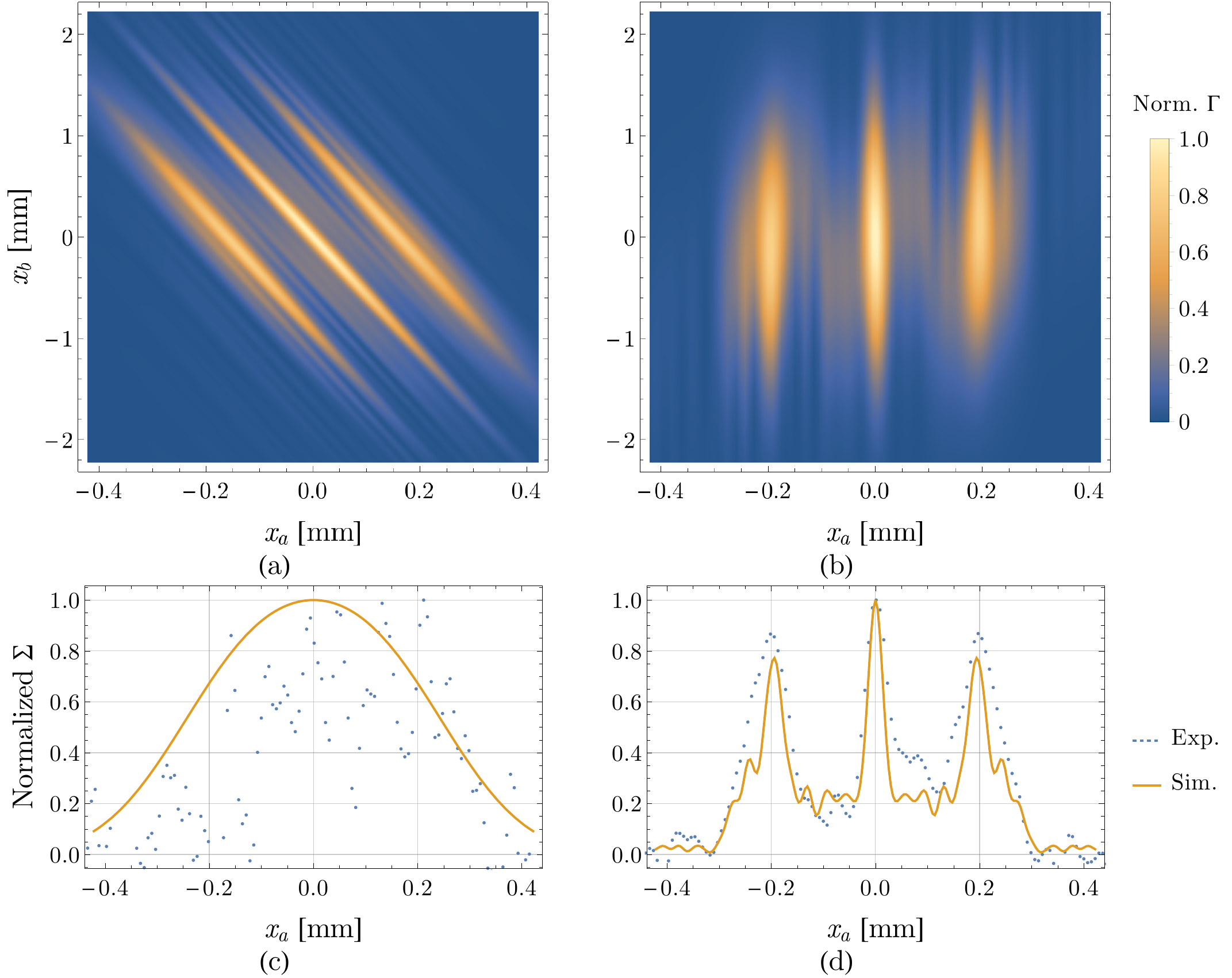}
\caption{(a) Simulation of CPI measurements obtained by evaluating pixel-by-pixel correlation between $S_a$ and $S_b$ in the same scenario employed to get the experimental results of Figure \ref{fig:exp}. (b) Result of the refocusing algorithm reported in the integrand of Eq.~(\ref{refocus}), as applied to the result of panel (a). (c) and (d): The solid lines are obtained by integrating the data of panel (a) and (b), respectively, over the angular sensor $S_b$. Figure (c) coincides with the standard ghost image, while (d) represents the refocused image of CPI, described by Eq.~(\ref{refocus}). The simulation is based on theoretical predictions reported in Appendix A-B, after integration over $y$ direction. Points are experimental data obtained by integrating over $y_a$ the experimental results of Figure \ref{fig:exp}. }
\label{fig:refocused}
\end{figure}

To understand the physical origin of the refocusing capability of CPI, we report in Figure \ref{fig:refocused}(a) the result of the pixel-by-pixel correlation of the intensity fluctuations evaluated on the planes of $S_a$ and $S_b$  [i.e., $ \Gamma_{z_a,z_b}(\bm{\rho}_a, \bm{\rho}_b)$ from Eq.\ (B1), after integration over $y_a$ and $y_b$], in the same experimental scenario of Figure \ref{fig:exp}. For each pixel of the angular sensor $S_b$, we observe on $S_a$ a displaced image of the object: Hence, imaging the light source on the high-resolution sensor $S_b$ enables changing the perspective on the observed scene. This result explains why the standard ghost image reported in Figure \ref{fig:refocused}(c) is blurred: When no angular information is retrieved (i.e., when integration over $S_b$ is performed), all displaced images combine into the out-of-focus image $\Sigma_{z_a,z_b} (\bm{\rho}_a) = \int d^2\bm{\rho}_b\,\Gamma_{z_a,z_b}(\bm{\rho}_a,\bm{\rho}_b)$. In ghost imaging, integration performed by the bucket detector clearly erases the precious information contained in the raw data of CPI. On the contrary, CPI exploits the extra information gained by the high resolution detector $S_b$. As shown in Figure \ref{fig:refocused}(b), all displaced images are realigned by the reshaping and resizing algorithm that appears in the integrand of Eq.~(\ref{refocus}), hence, no blurring occurs anymore upon integration over $S_b$, and the refocused image of Figure \ref{fig:refocused}(d) is obtained. Figures \ref{fig:refocused}(c) and (d) also show the excellent agreement between experimental data (points) and theoretical predictions (solid line).

\begin{figure}
\centering
\includegraphics[width=0.48\textwidth]{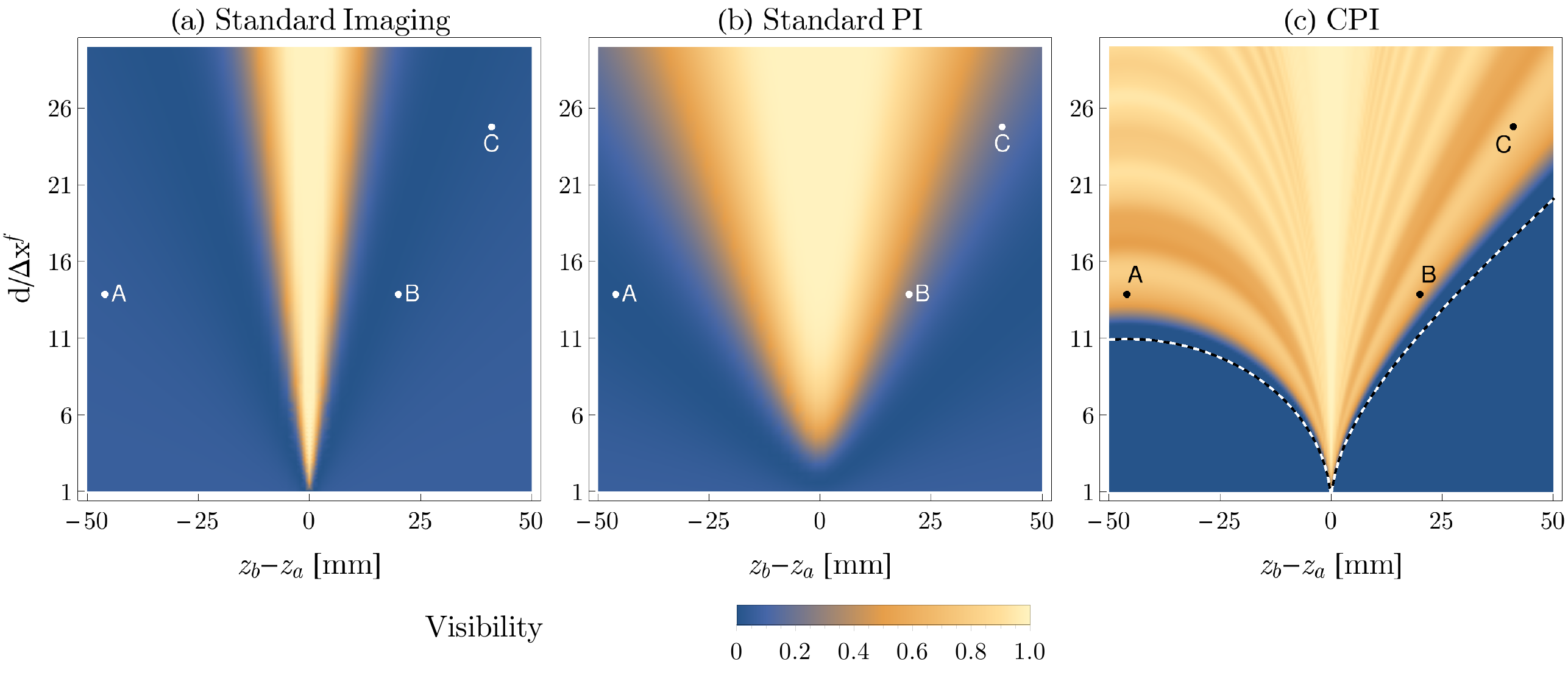}
\caption{Analysis of the range of perfect refocusing for double-slit objects with $d=2a$; the slit distance $d$ is normalized to the resolution of the focused image $\Delta x^f$. Visibility is computed by considering (a) standard imaging, (b) standard plenoptic imaging (with $N_u=3$), and (c) CPI devices sharing the same NA employed in the experiment. Points A, B and C correspond to the experimental scenarios leading to the results reported in Figure \ref{fig:exp} above and Figures 7(a) and (b); the (white) dashed line in panel (c) is the geometrical perfect refocusing limit given by Eq.~(\ref{DOF}).}\label{fig:visib}
\end{figure}

Let us now focus our attention on the central point of the paper, namely, the unique opportunity offered by CPI to refocus without sacrificing diffraction-limited image resolution, as defined by the numerical aperture of the imaging system. In Figure \ref{fig:visib}(c), the dashed (white) line represents geometrical-optics prediction for the maximum range of ``perfect'' refocusing in CPI, given by \cite{cpi_prl}
\begin{equation}\label{DOF}
\left| 1-\frac{z_a}{z_b} \right| < \frac{\Delta x}{\Delta u} =\frac{ d z_a/z_b  }{\max[\lambda z_b/a, 2 \lambda/(M_b \mathrm{NA}_{b}), 2 \delta u/M]}
\end{equation}
with $\Delta x$ the resolution on sensor $S_a$, and $\Delta u$ the resolution on the source plane. In the right hand side of Eq.~(\ref{DOF}), we have expressed both quantities in the simple case of double-slit objects of width $a$ and center-to-center distance $d=2a$.
The resolution $\Delta x=d z_a/z_b$ is defined by the geometrical projection of the image of the mask on the sensor plane. The resolution $\Delta u$ is defined by the larger contribution associated with diffraction at the object (i.e., $\lambda z_b/a$), numerical aperture of $L_b$ and pixel size $\delta_u$; these last two contributions enter into play for objects quite close to the light source [i.e., for $z_b =2 a/(M_b \mathrm{NA}_{b})$ and $z_b = 2 \delta_u a/(M\lambda)$, respectively]. Based on Eq.\ (\ref{DOF}), the physical quantities defining the spatial and the angular resolution of CPI are thus the object position $z_b$ and the object features $a$ and $d$. The density plot in Figure \ref{fig:visib}(c) reports visibility $V(d/\Delta x^f, z_b-z_a)$ of the refocused CPI images of double-slit masks, evaluated in the present experimental setup (see Appendix A-B). Besides giving the degree of reliability of the geometrical prediction of Eq.~(\ref{DOF}), this plot unveils the physical limit of resolution and DOF in CPI.

To compare CPI with both standard imaging and standard PI, we consider imaging devices having the same NA as the light source in our experiment, and report in Figures \ref{fig:visib}(a)-(b) the visibility they achieve. For standard PI, we have considered $N_u=3$ to avoid strongly compromising image resolution. Comparison of Figures \ref{fig:visib}(a), (b) and (c) indicates that CPI combines at best the advantages of standard and plenoptic imaging: It preserves the resolution of standard imaging while increasing the DOF even beyond the typical values of standard PI. Interestingly, for close up ($z_b<z_a$), object details larger than $d \gtrsim \sqrt{8\lambda z_a} \simeq 2.8 \Delta x^f$ (the refocusing limit corresponding to $z_b=z_a/2$) can always be refocused by CPI, no matter how close the object is to the source. For $z_b>z_a$, the maximum achievable depth of field is significantly larger than in both standard imaging and standard PI. As demonstrated in Appendix D, the refocusing range in CPI is limited by interference and diffraction at the object, for close-up, and only by diffraction, for distant objects. Such dependence can be understood in terms of Klyshko picture \cite{klyshko}, as applied to ghost imaging with chaotic light \cite{valencia}. Hence, CPI reaches the fundamental limits imposed by the wave nature of light to both image resolution and DOF. 

Points A, B, and C in Figure \ref{fig:visib} represent the experimental scenarios corresponding to the results reported in Figure \ref{fig:exp} (B), and in Figures 7(a)-(b) (A and C). In all three points CPI clearly guarantees a significant DOF advantage. In particular, the object corresponding to A and B can be refocused by CPI in a range more than $7$ times larger than in standard imaging, and $2.5$ larger than in a standard PI device characterized by a three times worst spatial resolution ($N_u=3$). For the wider object corresponding to point C, the maximum achievable DOF with CPI is $4$ times larger than with standard imaging, and twice larger than with a standard PI with $N_u=3$. It is worth emphasizing that the DOF of the standard ghost image represents the axial resolution of CPI ($\Delta z^{CPI}=\lambda / \mathrm{NA}^2$); hence, the ratio between the depth of fields of CPI and standard imaging fixes the number of planes that can be refocused by CPI.

\begin{figure}
\centering
\includegraphics[width=0.45\textwidth]{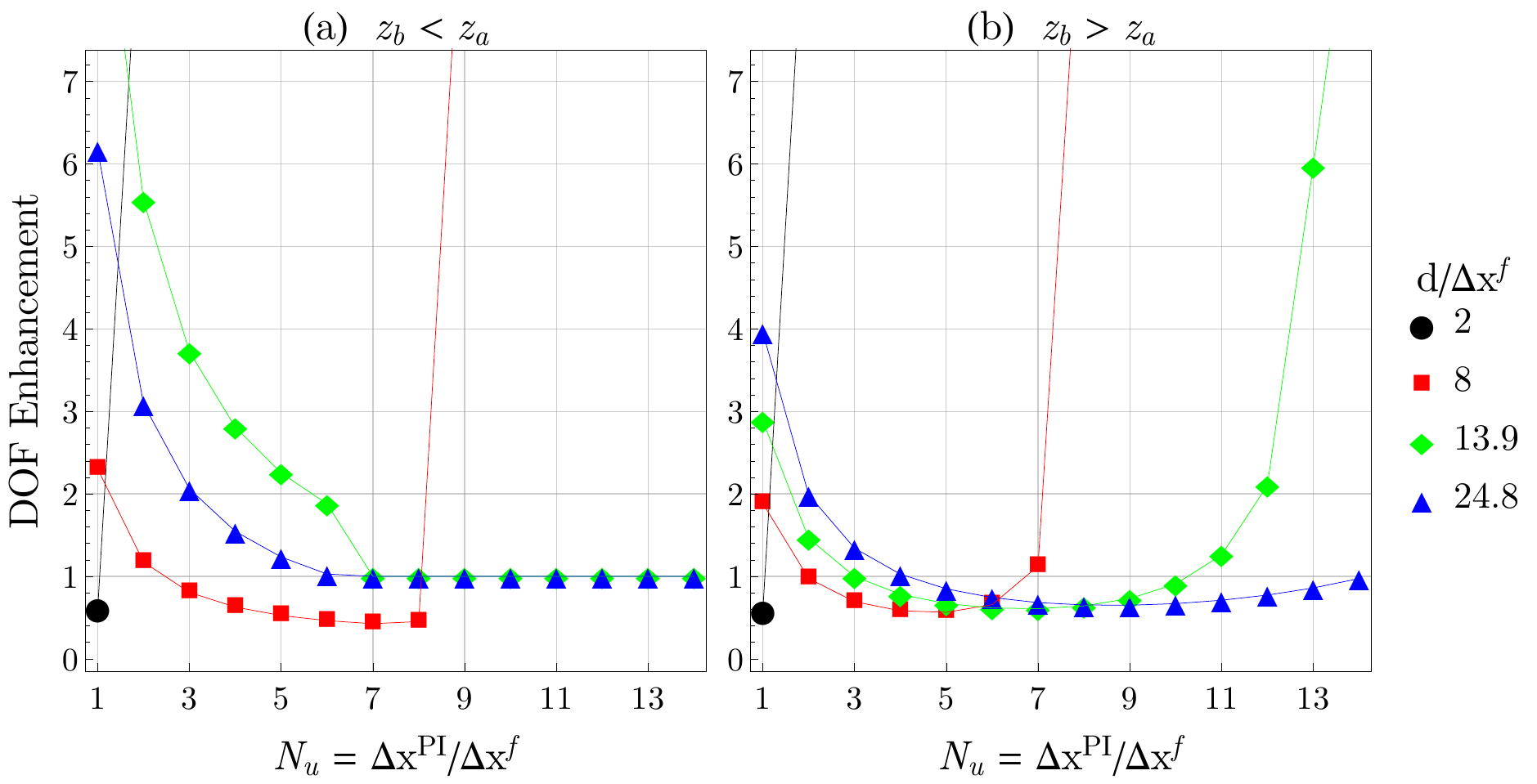}
\caption{Comparison between standard PI and CPI in terms of maximum achievable DOF versus resolution. The comparison is made for double-slit masks of varying distance $d$, and width $a=d/2$. DOFs are evaluated by considering the maximum ($z_b^M$) and minimum ($z_b^m$) values of the object distance for which the image is still resolved based on Rayleigh criterion (namely, $V\geq10\%$). We report the results for (a) $z_b<z_a$, and (b) $z_b>z_a$ obtained by considering the present experimental setup for CPI, and an equivalent standard PI device. 
}\label{fig:DOFvsRES}
\end{figure}

To summarize the above results, in Figure \ref{fig:DOFvsRES}, we plot the DOF enhancement offered by CPI with respect to standard PI as a function of the resolution compromise of conventional PI. The DOF enhancement is defined as the ratio between the maximum achievable DOF of CPI and standard PI; the resolution compromise of standard PI corresponds to the ratio between the maximum image resolutions of standard PI and CPI/standard imaging and is represented by the number of angular pixels $N_u$ of standard PI. All parameters are the same employed in Figure \ref{fig:visib}. To better emphasize the lack of refocusing limit for close up, we have chosen to separately plot the two cases of object closer to and farther away from the conjugate plane $z_b=z_a$.  In line with the results in Figure \ref{fig:visib}, CPI always outperforms standard PI. In fact, the DOF of CPI is generally larger than for standard PI, although there are ranges of $N_u$ where PI may overcome the DOF of CPI by loosing resolution.

\textit{Conclusions and outlook.---} We have demonstrated that CPI can push plenoptic imaging to its fundamental limits of both resolution and maximum achievable DOF: Unlike standard PI, CPI has no constraints on image resolution, which stays diffraction-limited as in standard imaging systems. Still, CPI enables increasing the DOF well beyond the typical value of standard imaging. The advantages of both standard and plenoptic imaging are thus combined at best in CPI, whose maximum achievable DOF is solely limited by interference and diffraction at the object (see, e.g., Figures 9-10). Several technologies have been introduced in the past years where light correlation properties enable going beyond the capabilities of standard imaging systems (e.g., \cite{pittman,valencia,qu_superres,qu_superres2,sofi,undetected}); however, in most cases, previous technologies exploited the correlations in either position or momentum, but not both. The simultaneous use of both momentum and position correlation has so far only been used for fundamental demonstrations \cite{bennink,dangelo_kim,scarcelli_er}; here, for the first time, it is exploited to push the fundamental limits of practical imaging systems.

CPI has the potential to strongly improve the performances of both microscopy, where high lateral and axial resolutions are required together with large DOF, and 3D imaging, where fast multiperspective acquisitions are desired. Future studies will be devoted to acquisition time optimization, considering hardware (fast CMOS, smart sensors \cite{remondino2013tof}) and software solutions (compressed-sensing and sparse measurement techniques \cite{katz}) to regain the single-shot advantage of conventional plenoptic imaging.

\begin{acknowledgments}
The Authors thank T. Macchia, C. Plantamura and and C. Bevilacqua for participating at the preliminary experimental activity. MD acknowledges financial support from the Italian Ministry of Education, University and Research (MIUR), projects PONa3\_00369 (``Laboratorio per lo Sviluppo Integrato delle Scienze e delle Tecnologie dei Materiali Avanzati e per dispositivi innovativi -LABORATORIO SISTEMA''). MD, FD, AM and AG acknowledge financial support from MIUR, project n.~PON02-00576-3333585 (P.O.N. RICERCA E COMPETITIVIT\`A 2007-2013 - Avviso n.~713/Ric. del 29/10/2010, Titolo II - ``Sviluppo/Potenziamento di DAT e di LPP''). MD, FD, AG and FVP are partially supported by Istituto Nazionale di Fisica Nucleare (INFN) through the projects ``QUANTUM'' and ``PICS''.
\end{acknowledgments}

\appendix

\section{CPI with chaotic light}

Let us review the main steps for demonstrating the refocusing capability of CPI, and Eq.~(1) of the main text; a complete discussion can be found in Refs.~\cite{cpi_prl,cpi_qmqm}. Thank to the chaotic nature of our light source, the intensities measured at point $\bm{\rho}_a$ of the sensor $S_a$ at time $t_a$ and at point $\bm{\rho}_b$ of the sensor $S_b$ at time $t_b$ are characterized by spatio-temporal correlations, described by the Glauber correlation function \cite{scully}
\begin{eqnarray}\label{glauber}
G^{(2)}(\bm{\rho}_a,\bm{\rho}_b;t_a,t_b)\! & \!=\! & \!\left\langle E^{(-)}_a
(\bm{\rho}_a,t_a) E^{(-)}_b (\bm{\rho}_b,t_b) \right. \nonumber
\\ & & \left. E^{(+)}_b (\bm{\rho}_b,t_b) E^{(+)}_a
(\bm{\rho}_a,t_a) \right\rangle.
\end{eqnarray}
Here, $E_{i}^{(\pm)}$ are the positive- and negative-frequency components of the electric field [$E^{(+)}=(E^{(-)})^{\dagger}$] at each detector, and the expectation value is evaluated by considering the quantum state $\varrho$ of the source. The fields are propagated from the output plane of the source to each detector by the paraxial optical transfer functions $g_a$ and $g_b$, as
\begin{equation}\label{field}
E^{(+)}_i (\bm{\rho}_i,t_i) = \int d\Omega \int d^2\bm{\kappa}\, a_{\bm{k}} e^{-i\Omega t_i} g_i(\bm{\rho}_i,\bm{k}),
\end{equation}
where the integral runs over frequencies $\Omega$ and transverse momenta $\bm{\kappa}$, and $a_{\bm{k}}$ is the canonical field operator of the mode $\bm{k}$. In the paraxial approximation, the 3D wave vector reads $\bm{k}=(\bm{\kappa},\Omega/c)$. When the source is both stationary and quasi-monochromatic, with peak frequency $\omega$, the correlation function depends only on $\tau=t_a-t_b$, and the time dependent part approximately factorizes with respect to the space-dependent part. In addition, for a chaotic source, the four-point expectation value involved in Eq.\ (\ref{glauber}) reduces to the sum of two terms,
\begin{eqnarray}
\langle a_{\bm{k}_1}^{\dagger} a_{\bm{k}_2}^{\dagger} a_{\bm{k}_3} a_{\bm{k}_4} \rangle & \propto & \delta (\bm{k}_1-\bm{k}_4) \delta (\bm{k}_2-\bm{k}_3) \nonumber \\ & & + \delta (\bm{k}_1-\bm{k}_3) \delta (\bm{k}_2-\bm{k}_4).
\end{eqnarray} 
Therefore, upon neglecting the time dependence (i.e., ~working within the coherence time of the source), the second-order correlation function in Eq. (\ref{glauber}) reads
\begin{equation}\label{G2thermal}
G^{(2)}(\bm{\rho}_a,\bm{\rho}_b) = I_a (\bm{\rho}_a) I_b (\bm{\rho}_b) + \Gamma(\bm{\rho}_a,\bm{\rho}_b),
\end{equation} 
where the first term is the mere product of intensities at the points $\bm{\rho}_i$ on $S_i$, with $i=a,b$. The second term
\begin{equation}\label{Gamma}
\Gamma(\bm{\rho}_a,\bm{\rho}_b) = \left| \int d^2\bm{\kappa} g_a^*(\bm{\rho}_a,\bm{\kappa}) g_b(\bm{\rho}_b,\bm{\kappa}) \right|^2
\end{equation}
represents the nontrivial part of the second-order correlation, that yields the correlation of intensity fluctuations and encodes plenoptic imaging properties. 

To unveil such properties, we first need to compute the transfer functions in the setup of Figure~1 of the main text. Up to irrelevant normalization factors and phases, we have:
\begin{equation}\label{ga}
g_a(\bm{\rho}_a,\bm{\kappa}) = \int \! d^2\bm{\rho}_s f(\bm{\rho}_s) \e^{\i\left(\bm{\kappa} - \frac{\omega}{cz_a}\bm{\rho}_a \right)\cdot\bm{\rho}_s} G(\bm{\rho}_s)_{\left[\frac{\omega}{cz_a}\right]}
\end{equation}
in the reflected arm and
\begin{eqnarray}\label{gb}
g_b(\bm{\rho}_b,\bm{\kappa}) & = & \int\! d^2\bm{\rho}_s\!\int\! d^2\bm{\rho}_o f(\bm{\rho}_s) A (\bm{\rho}_o) \nonumber \\ & & \times G(\bm{\rho}_s)_{\left[\frac{\omega}{cz_b}\right]} \e^{\i\bm{\kappa}\cdot\bm{\rho}_s - \frac{\i\omega}{cz_b} \bm{\rho}_o\cdot \left( \bm{\rho}_s + \frac{\bm{\rho}_b}{M} \right)} 
\end{eqnarray}
in the transmitted arm, with $G(\bm{\rho})_{[\beta]}= \exp(i\beta\bm{\rho}^2/2)$, $A(\bm{\rho}_o)$ the object transmission function, and $f(\bm{\rho}_s)$ the source amplitude profile. Notice that in Eq.~(\ref{gb}) we have assumed that the lens $L$, that focuses the image of the source on the sensor $S_b$ with magnification $M$, is diffraction-limited. Given the propagators of Eqs.~(\ref{ga})-(\ref{gb}), one can compute the nontrivial part of the correlation function in Eq.~(\ref{Gamma}), which yields, up to trivial factors,
\begin{eqnarray}\label{Gammazazb}
& &\Gamma_{z_a,z_b} (\bm{\rho}_a,\bm{\rho}_b) = \Biggl|\!\int\! d^2\bm{\rho}_o\!\int\! d^2\bm{\rho}_s A(\bm{\rho}_0) F(\bm{\rho}_s)  \nonumber \\
& & \quad \times G(\bm{\rho}_s)_{\left[\frac{\omega}{c}\!\left(\frac{1}{z_b}- \frac{1}{z_a} \right)\!\right]} \e^{-\frac{\i\omega}{c z_b} \!\left[\left( \bm{\rho}_o - \frac{z_b}{z_a} \bm{\rho}_a \right)\! \cdot \bm{\rho}_s + \bm{\rho}_o \cdot \frac{\bm{\rho}_b}{M} \right]} \Biggr|^2,
\end{eqnarray}
where $F=|f|^2$ is the source intensity profile. 

On one hand, Eq.~(\ref{Gammazazb}) indicates that a focused {\it coherent image} of the object $A(\bm{\rho}_0)$ is obtained  when $z_a=z_b$. The focusing condition is shared by the ghost image, which is given by the integral of Eq. (\ref{Gammazazb}) over the sensor $S_b$: 
\begin{eqnarray}\label{image_def}
\Sigma_{z_a,z_b} (\bm{\rho}_a) = \!\int\! d^2\bm{\rho}_b \Gamma_{z_a,z_b} (\bm{\rho}_a,\bm{\rho}_b) .
\end{eqnarray}
Interestingly, the focused ghost image
\begin{equation}\label{image}
\Sigma_{z_a,z_a} (\bm{\rho}_a) \propto \!\int \! d^2\bm{\rho}_o |A(\bm{\rho}_o)|^2 \left| \tilde{F} \left[ \frac{\omega}{cz_a} \left( \bm{\rho}_o - \bm{\rho}_a \right) \right] \right|^2 
\end{equation}
is formally identical to a standard {\it incoherent image} (with no magnification): It sets a quasi one-to-one correspondence between points of the object plane and points of the sensor $S_a$. The point-spread function (PSF) of the ghost image is given by the squared modulus of the Fourier Transform of the source intensity profile ($\tilde{F}$); in ghost imaging, the source intensity profile thus plays the exact same role that a lens plays in standard imaging. The image resolution $\Delta\rho_a \simeq 2\pi c z_a/(\omega D_s) =: \lambda/NA$ is defined by the numerical aperture NA of the focusing element (here, the source), characterized by the effective diameter $D_s$. 

On the other hand, due to the first-order image of the source on sensor $\mathrm{D}_b$, Eq.~(\ref{Gammazazb}) also entails a correspondence between points of the source plane and pixels of the sensor $S_b$ ($\bm{\rho}_b=-M\bm{\rho}_s$), whose uncertainty $\Delta\rho_b = M \lambda z_b/a$ is determined by the typical size $a$ of the smallest detail of the object, that acts as a pupil for the lens $L$, and by the distance $z_b$. The resolution of the source image is thus limited by diffraction at the object.

The plenoptic properties of $\Gamma_{(z_a,z_b)}$ clearly emerge in the geometrical optics limit $\omega\to\infty$. Indeed, in this limit, the double integral in Eq.~(\ref{Gammazazb}) can be approximated by a stationary-phase approximation, which reveals that the object and source points that provide the prominent contribution to the integral are related to the detection points by
\begin{eqnarray}
\bm{\rho}_s & = & - \frac{\bm{\rho}_b}{M}, \\
\bm{\rho}_o & = & \frac{z_b}{z_a} \bm{\rho}_a - \frac{\bm{\rho}_b}{M} \left( 1-\frac{z_b}{z_a} \right),\label{eq:refocusing}
\end{eqnarray}
The first result is expected by the aforementioned first-order imaging of the source. The second line is nontrivial, since it is not related to any first-order imaging property, but it connects points of the object to points of \textit{both} sensors. Thus, in the geometrical optics limit, we can deduce the asymptotic behavior of the nontrivial part of the second-order correlation function:
\begin{equation}\label{scaling0}
\Gamma_{z_a,z_b} (\bm{\rho}_a,\bm{\rho}_b) \sim F\!\left( - \frac{\bm{\rho}_b}{M} \right)^2 \!\left| A\!\left[ \frac{z_b}{z_a} \bm{\rho}_a - \frac{\bm{\rho}_b}{M} \left( 1-\frac{z_b}{z_a} \right) \right] \right|^2 .
\end{equation}
If $z_b\neq z_a$, the integration of this result over ${\bf \rho}_b$, which is equivalent to retrieving an out-of-focus (incoherent) ghost image, erases information on the aperture function of the object, and leads to a blurred image (see Figure 3(c) of the main text). This indicates the crucial role played by the high-resolution detector $S_b$, as opposed to the bucket detector of standard ghost imaging. In fact, based on Eq.~(\ref{scaling0}), one can use the information obtained from intensity correlation measurements to \textit{refocus} the coherent correlation plenoptic image of the object, by applying a refocusing algorithm that is very similar to the one of standard PI \cite{ng}:
\begin{equation}\label{scaling}
\Gamma_{z_a,z_b} \left[\frac{z_a}{z_b} \bm{\rho}_a - \frac{\bm{\rho}_b}{M} \left( 1-\frac{z_a}{z_b} \right),\bm{\rho}_b\right] \sim F\!\left( - \frac{\bm{\rho}_b}{M} \right)^2 \!\left| A(\bm{\rho}_a) \right|^2 .
\end{equation}
The integration of the refocused correlation function of Eq.~(\ref{scaling}) over $\bm{\rho}_b$ [as reported in Eq.~(1) of the main text] increases the signal-to-noise ratio of the final refocused image obtained by CPI without blurring it, as demonstrated in Figure~3(d) of the main text.

\section{CPI with a Gaussian source}

\noindent The case of a source with Gaussian intensity profile $F(\bm{\rho}_s) = \exp [ - \bm{\rho}_s^2/(2 \sigma^2) ]/ (2\pi\sigma^2)$ is particularly relevant, since it is analytically feasible,and, most important, can be used to model our experimental source. 
Based on Eq.~(\ref{Gammazazb}), the correlation of intensity fluctuations can be expressed as the convolution of the aperture function of the object with the (coherent) CPI point-spread function (PSF) $\mathcal{C}(\bm{\rho})$, namely,
\begin{equation}\label{Gamma_bis}
\Gamma_{z_a,z_b} (\bm{\rho}_a,\bm{\rho}_b)\!=\!\left| \!\int d^2\bm{\rho}_o A(\bm{\rho}_o) \e^{-\frac{\i\omega \bm{\rho}_o \cdot \bm{\rho}_b }{c z_b M} } \mathcal{C}\!\left(\bm{\rho_o}-\frac{z_b}{z_a} \bm{\rho}_a\right) \!\right|^2,
\end{equation}
where, for a Gaussian source \cite{cpi_qmqm},
\begin{equation}\label{coh2}
\mathcal{C}(\bm{\rho}) \propto \exp\!\left( -\frac{1}{2} \!\left( \frac{\omega\sigma}{c z_b} \right)^2\! \frac{|\bm{\rho}|^2}{1 - \frac{\i\omega\sigma^2}{cz_b}\left(1-\frac{z_b}{z_a}\right)} \right).
\end{equation}
It is also immediate to verify that the PSF $\mathcal{J}$ of the (incoherent) ghost image described by Eq.~(\ref{image_def}), is related to the PSF of CPI by the relationship $\mathcal{J}(\bm{\rho})\propto |\mathcal{C}(\bm{\rho})|^2$. In the case of a Gaussian source, we have \cite{cpi_qmqm}:
\begin{equation}\label{inc2}
\mathcal{J}(\bm{\rho}) \propto \exp\!\left( - \!\left( \frac{\omega\sigma}{c z_b} \right)^2\! \frac{|\bm{\rho}|^2}{1 + \left(\frac{\omega\sigma^2}{cz_b} \left( 1-\frac{z_b}{z_a}\right) \right)^2} \right).
\end{equation}

The result in Eq.~(\ref{inc2}), typical of both ghost imaging and conventional incoherent imaging, also enables a clearer understanding of Figure~4 of the main text, where the visibility of the incoherent image is asymmetric with respect to the object position $z_b=z_a$. In fact, the asymmetry can be appreciated only in panel (c) of Figure~4 of the main text, and would disappear if one fixes the object distance $z_b$ and varies the sensor distance $z_a$. Indeed, when $z_a$ is kept fixed as in our experimental case, the image of a point object placed in $\bm{\rho}_o$, at a distance $z_b\neq z_a$ from the source, is a Gaussian centered in $\bm{\rho}_a = \bm{\rho}_o /\alpha$, with $\alpha=z_b/z_a$, and characterized by the width
\begin{equation}\label{sigmai}
\frac{\sigma_i (\alpha)}{\alpha} = \frac{1}{\alpha} \sqrt{ \frac{1}{2} \!\left( \frac{c z_a}{\omega\sigma} \right)^2\! \alpha^2 + \frac{\sigma^2}{2}\left(1-\alpha\right)^2 }.
\end{equation} 
The factor $1/\alpha$ is a magnification due to the geometrical projection of the object onto a plane different from its conjugate image plane. Hence, the quantity that actually determines the image resolution in Eq.~(\ref{sigmai}) is $\sigma_i(\alpha)$, and is minimized by
\begin{equation}
\bar{\alpha}= \!\left( 1 + \!\left( \frac{c z_a}{\omega\sigma^2} \right)^2 \right)^{-1}. 
\end{equation}\label{alphamin}
The quantity $\bar{\alpha}$ is always smaller than one, hence, the image resolution is minimum for $\bar{z}_b=\bar{\alpha} z_a < z_a$, with $\bar{\alpha}$getting smaller for larger source widths, as well as smaller wavelengths and $z_a$.

\section{Experimental methods}

\begin{figure}
\centering
\includegraphics[width=0.48\textwidth]{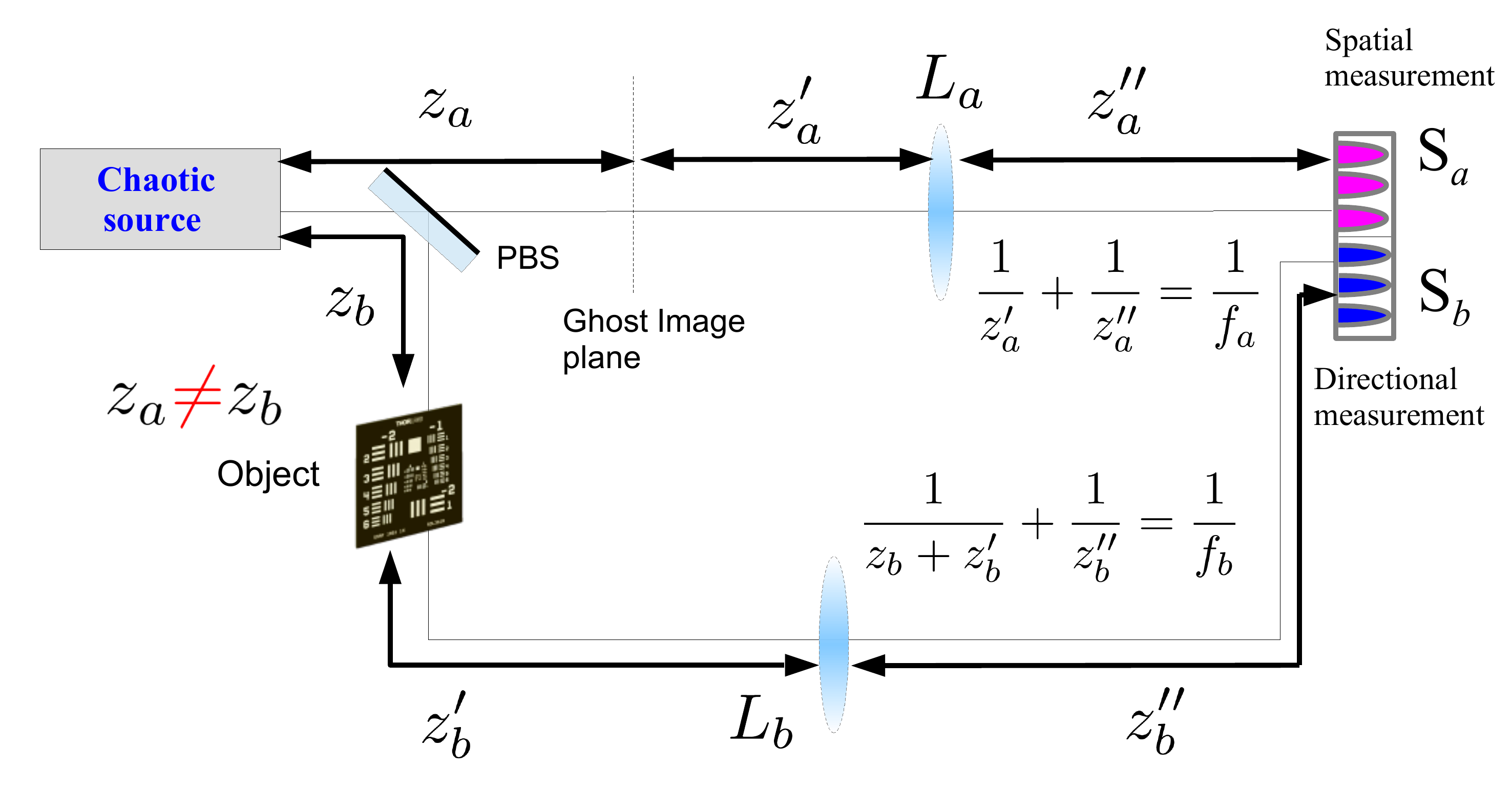}
\caption{Experimental setup employed for demonstrating correlation plenoptic imaging. The spatial and angular sensors $S_a$ and $S_b$, respectively, are different regions of a sCMOS camera; this has inforced the use of an additional lens $L_a$ to reproduce the ghost image plane on the sensor $S_a$.}\label{fig:setup_camera}
\end{figure}

The detailed experimental setup is reported in Figure \ref{fig:setup_camera}.

The chaotic source is made by a CW single-mode laser with wavelength $\lambda = 532\,\mathrm{nm}$ and tunable power up to $5 W$ (Azur Light Systems ALS-532nm-SF). The laser beam, after being expanded to give a spot size of $\sigma=1.08\:\mathrm{mm}$, passes through a polarizer before impinging on a rotating ground glass disk, spinning at $0.05\:\mathrm{Hz}$, at a distance of about $4\:\mathrm{cm}$ from the center. Light from the source is then divided by a polarizing beam-splitter (PBS); the combination of polarizer and PBS enables balancing the intensities at the sensors $S_a$ and $S_b$, which are different regions of the same sCMOS camera (Hamamatsu ORCA-Flash 2.8 camera C11440-10C), thus maximizing their SNR. The reflected beam passes through the object of interest (ThorLabs 1951 USAF Resolution Test Targets), propagates toward a lens ($L_b$) of focal length $f_b=300\, \mathrm{mm}$, and reaches the angular sensor $S_b$. The angular detector $S_b$ is in the conjugate plane of the source, whose image is magnified by $M_b=1$. The transmitted beam propagates toward a lens ($L_a$)  of focal length $f_a=125\,mm$, reproducing on the spatial sensor $S_a$, with a magnification $M_a = 1$, the image of the ghost imaging plane (set at a distance $z_a=92\,	\mathrm{mm}$ from the source). The camera is characterized by a pixel size of $3.6\:\mathrm{\mu m}$, which is much smaller than both the spatial resolution (given by the diffraction limit associated with the smaller one between the source and the lens $L_a$ numerical apertures) and the angular resolution (given, for the chosen values of $z_b$ and object size, by diffraction at the object). We have thus performed a binning of the camera pixels to match the effective pixels $\delta x$ and $\delta u$ with the resolution of the spatial and angular measurements, respectively. In particular, during data acquisition, we have performed a 2x2 binning, to get $\delta_x=7.2\:\mathrm{\mu m} \approx \Delta x^f/2$, with $\Delta x^f= 14 \:\mu\mathrm{m}$. In post-processing, a further $10\times 1$0 binning was performed on the region of the camera sensor dedicated to the angular measurement, thus getting $\delta u=72\:\mathrm{\mu m} < \Delta u/2$, with $\Delta u=\lambda z_b/a$ in our experiment.

We acquire $50\,000$ frames for all measurements $A$, $B$ and $C$ at a frame rate of $45.4 \: \mathrm{s}^{-1}$, and with an exposition time $\tau_{\mathrm{meas}}=21\,\mathrm{\mu s}$ (the minimum enabled by the camera) approximately 100 times smaller than the source coherence time.
The acquired  frames are processed to evaluate the spatio-temporal correlation, which is expected to give the result of Eq.~(\ref{Gammazazb}), and the refocusing algorithm [as described by Eq.~(1) of the main text], thus getting the CPI images reported in both Figure 2 of the main text and in Figures 2 (a) and (b).
Indeed, in Figure \ref{fig:exp}(a) and (b), we report the experimental refocused images obtained, respectively, for element 3 of group 2 (having $d = 0.198 ~\mathrm{mm}$) placed at a distance $z_b=46\,\mathrm{mm}$ from the source (measurement A), and for element 4 of group 1 (having $d = 0.354 \,\mathrm{mm}$) placed at $z_b=133\,\mathrm{mm}$ (measurement C). Such experiments have been performed around the maximal displacements $z_b \neq z_a$ of the objects that were enabled by our setup. The SNR in Figure \ref{fig:exp}(a) is lower because that the displaced coherent images retrieved for $z_b=z_a/2$ are twice larger than the object, and distributed over a region wider than the illuminated area; the coherent images are thus often affected by a poor SNR, which reflects on the final refocused image. To avoid this issue, the divergence of the light source needs to be designed to account for such displacement and enlargement.

\begin{figure}
\centering 
\includegraphics[width=0.5\textwidth]{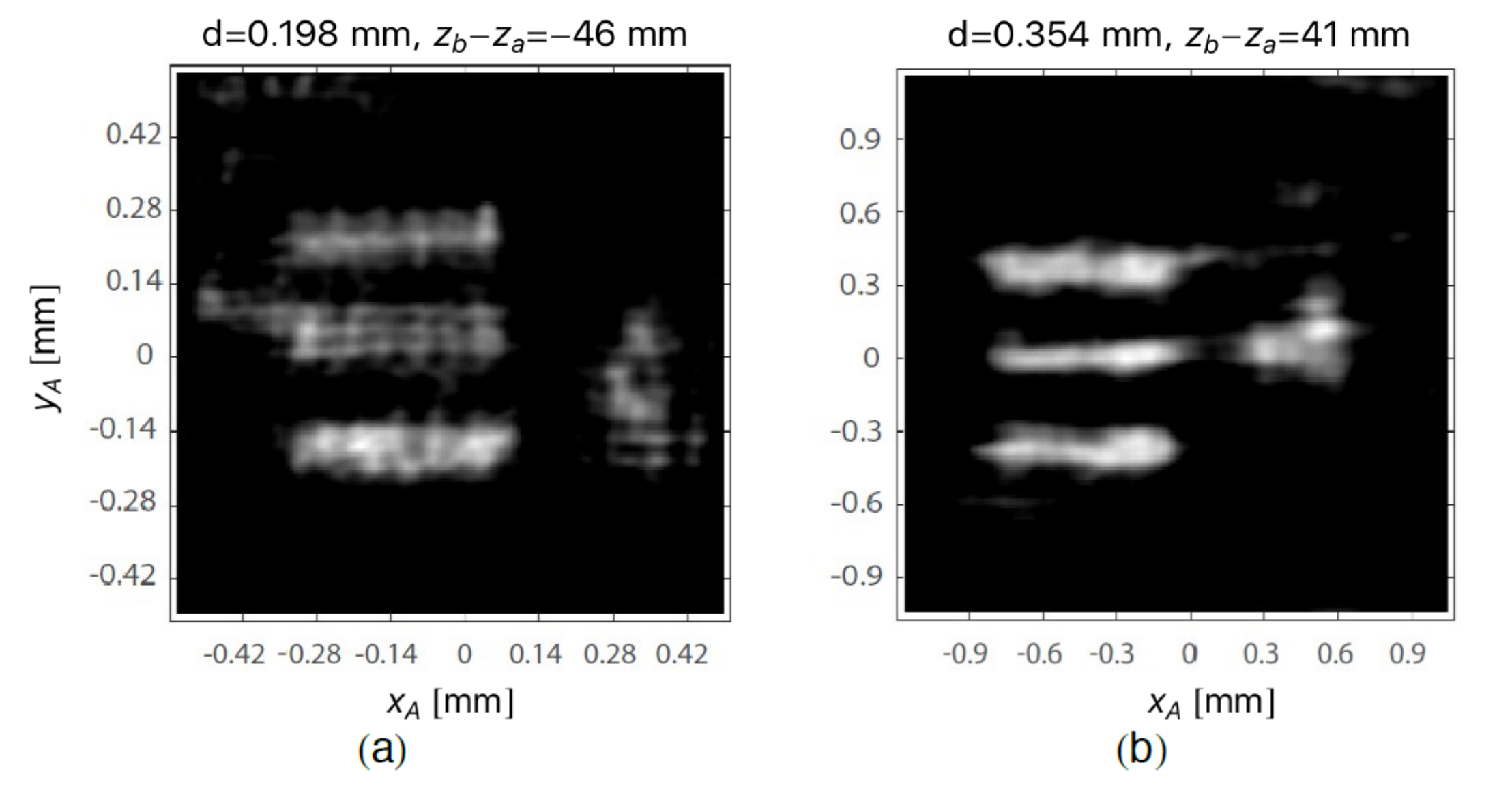}
\caption{Refocused images obtained in the experimental measurements A and C, as indicated in Figure 4 of the main text. The experimental data are taken with a pixel size at the diffraction limit ($\delta x = 7.2\:\mu\mathrm{m}$), while the refocused images have a pixel size scaled by a factor $z_b/z_a$, in line with Eq.~(12). After correlation measurement, low-pass Gaussian filtering and thersholding in the Fourier domain was applied to remove uncorrelated background.}\label{fig:exp}
\end{figure}

\section{DOF advantage of CPI}

\begin{figure}
\centering 
\includegraphics[width=0.47\textwidth]{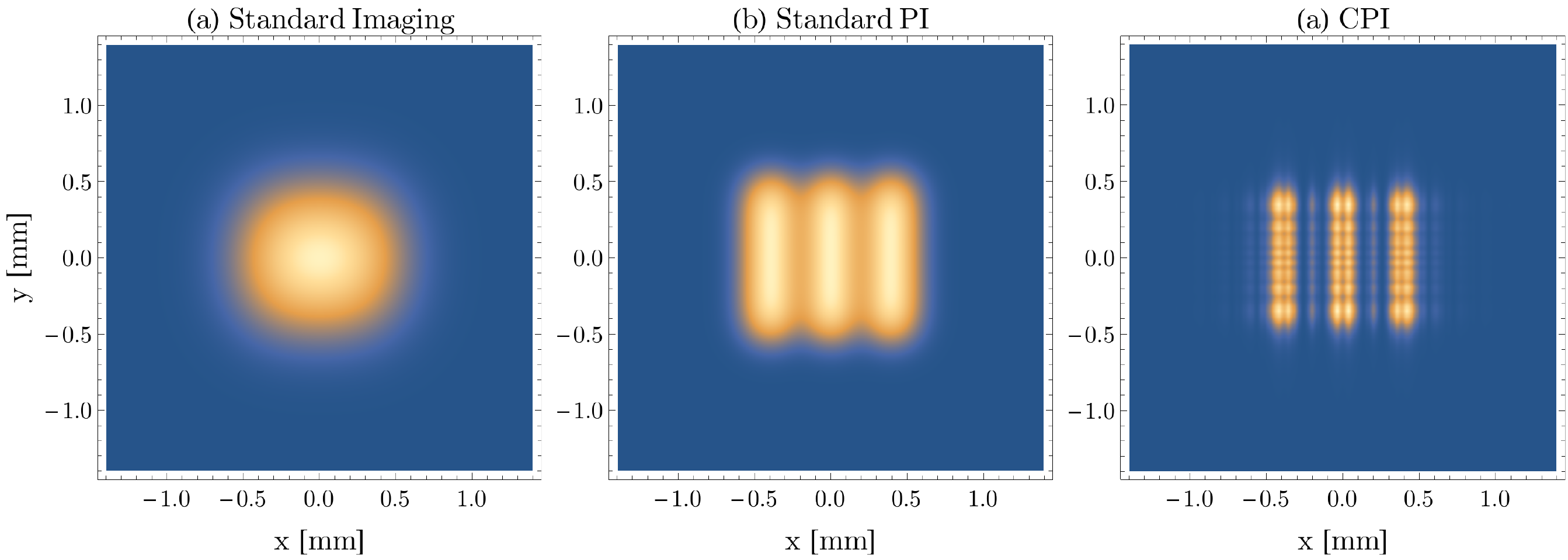}
\caption{Theoretical comparison between: (a) standard imaging, (b) standard plenoptic imaging (with $N_u=3$), and (c) CPI, for the same triple-slit of the measurement C. Both conventional imaging systems share the same NA as our setup.}
 \label{fig:comparison}
\end{figure}

In Fig. \ref{fig:comparison}, we report the comparison of CPI with both conventional imaging and conventional PI, in the same scenario corresponding to the experimental measurement (C) reported in Fig. \ref{fig:exp}(b): Beside gaining diffraction limited resolution, CPI clearly enables improving the DOF with respect to both conventional modalities.

\begin{figure}
\centering
\includegraphics[width=0.48\textwidth]{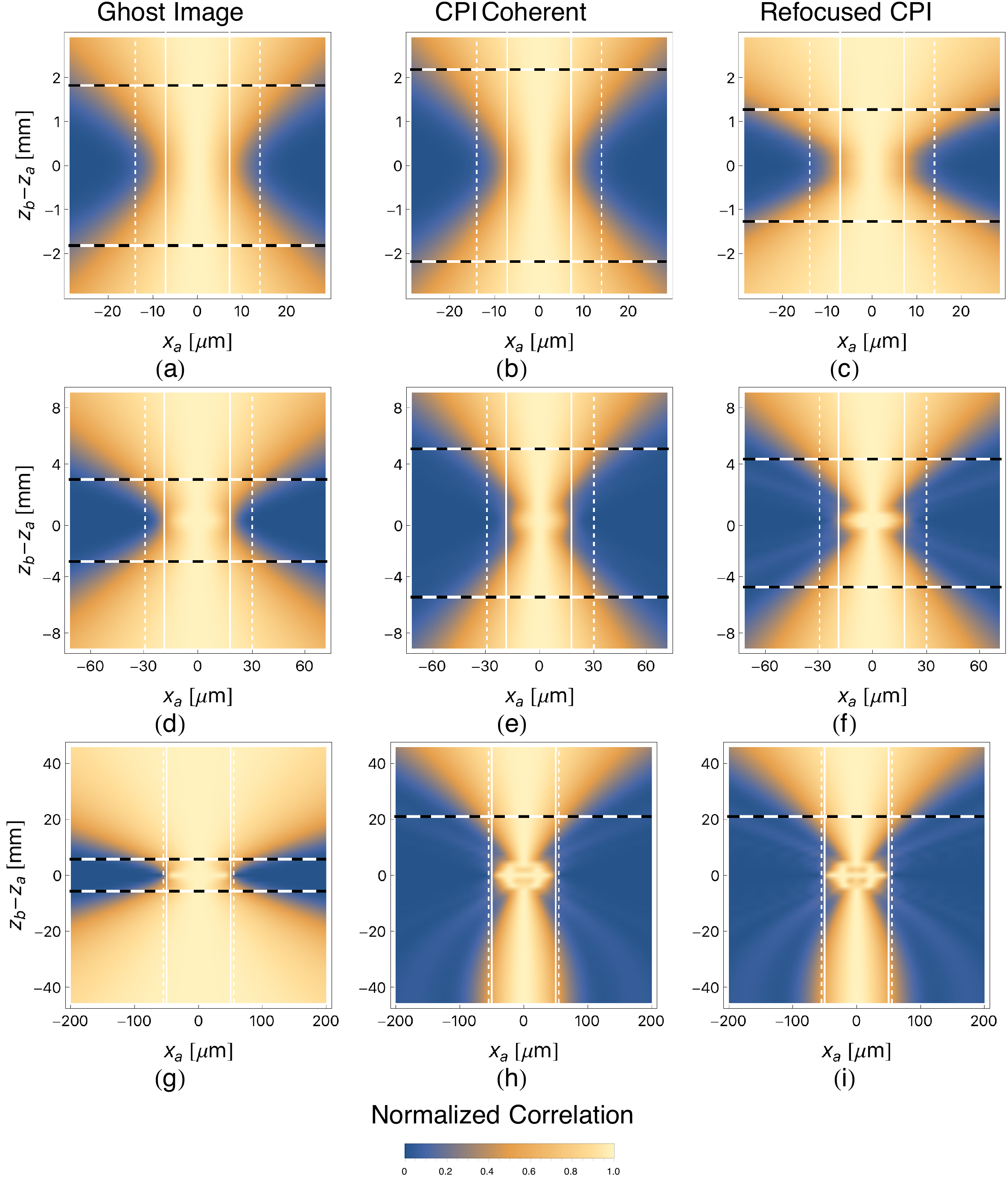}
\caption{Comparison between the incoherent ghost images [panels (a), (d), (g)], the coherent images from CPI [(b), (e), (h)], and the refocused image from CPI [(c), (f), (i)], for three different single-slit masks of width: $a=14\:\mu\mathrm{m}=\Delta x^f$ (top panels), $a=36\:\mu\mathrm{m}\simeq 2.5\:\Delta x^f$ (central panels), and $a=99\:\mu\mathrm{m}\simeq 7.2\:\Delta x^f$ (bottom panel), as for measurements A and B. The density plots report the correlation functions of Eq.s~(\ref{S_inc}), (\ref{Gamma_coh}), and (\ref{S_coh}), normalized to their peak value, evaluated in the experimental setup. The solid (white) lines represent the size of the object, while the (white) dashed lines represent the tolerance on the blurring of the images, namely the resolution limit. The (black) dotted lines indicate the DOF (see text for details).}\label{images-z}
\end{figure}

\begin{figure}
\centering
\includegraphics[width=0.5\textwidth]{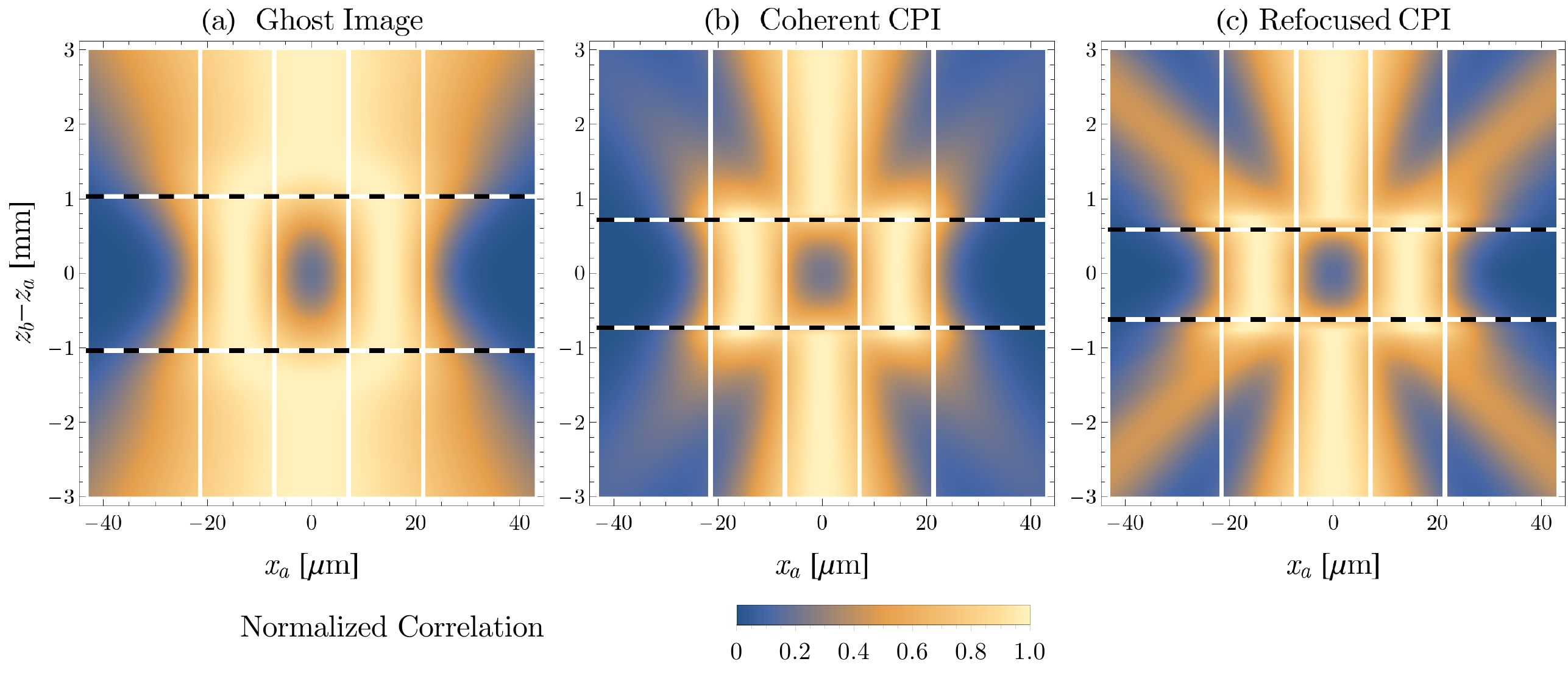}
\caption{Comparison between: (a) the incoherent ghost image, (b) the coherent image from CPI, and (c) the refocused image from CPI, for a double-slit mask of width $a=14\:\mu\mathrm{m}=\Delta x^f$ and slit separation $d=2a$. The density plots report the correlation functions of Eq.~(22)--(24), normalized to their value in $x_a = 0$, for any value of $z_b-z_a$, evaluated in the setup of Figure~1 of the main text. The solid (white) lines represent the size of the object. The  (black) dotted lines indicate the DOF, defined as the value of $z_b-z_a$ where the visibility drops below $10\%$ (see text for details).}
\label{double-slit}
\end{figure}

To analyze the fundamental limitations characterizing our CPI scheme, we shall compare the DOF of CPI and ghost imaging (which is always representative of standard imaging). In Figures \ref{images-z} and \ref{double-slit}, we report the incoherent ghost image (left column)
\begin{equation}
\Sigma_{z_a,z_b} \left( \frac{z_a}{z_b} x_a \right) = \int d x_o |A(x_o)|^2 \mathcal{J}_1 (x_o- x_a),
\end{equation}\label{S_inc}
the coherent image from CPI (central column)
\begin{equation}
\Gamma_{z_a,z_b} \left( \frac{z_a}{z_b} x_a, x_b=0 \right) = \left| \int d x_o A(x_o) \mathcal{C}_1 (x_o- x_a) \right|^2 ,
\end{equation}\label{Gamma_coh}
and the refocused image from CPI (rigth column)
\begin{equation}\label{S_coh}
\Sigma^{\mathrm{ref}}_{z_a,z_b} \left(x_a \right) = \int dx_b \Gamma_{z_a,z_b} \left( \frac{z_a}{z_b} x_a - \frac{x_b}{M} \left( 1- \frac{z_a}{z_b} \right) , x_b \right),
\end{equation}
with $\mathcal{J}_1 (x) = \int dy \mathcal{J}(x,y)$ and $\mathcal{C}_1 (x) = \int dy \mathcal{C}(x,y)$, where $\mathcal{J}$ and $\mathcal{C}$ are defined in Eqs.\ (\ref{coh2})-(\ref{inc2}). The results are shown for three single-slits of different width $a$, in Figure \ref{images-z}, and for a double-slit mask, in Figure \ref{double-slit}. The results have been obtained by considering the experimental setup of Figure~\ref{fig:setup_camera}, with a Gaussian source of width $\sigma=1.08$ mm, as retrieved from a fit of the measured source intensity profile. In line with the experiment, we fix $z_a$ while changing $z_b$.

In Figure~\ref{images-z}, the resolution limit (vertical dashed lines) is defined by adding to the half width $a/2$ of the object the quantity $2\sqrt{2 \ln 2}\sigma_{PSF}$ corresponding to the Rayleigh criterion for a Gaussian PSF of width $\sigma_{PSF}$; the DOF (horizontal dashed lines) is given by the value of $z_b-z_a$ at which the half width at half maximum of the image reaches the resolution limit. Based on both Figure~\ref{images-z}, the coherent image always has a wider DOF than ghost imaging. But what limits the maximum achievable DOF of the refocused correlation plenoptic image? On one hand, for close and wide enough objects ($z_b<z_a$ and $a \gtrsim 5 \Delta x^f$), Figure~\ref{images-z}(i) predicts an unlimited DOF for the coherent image, which results in an unlimited refocusing capability for CPI. On the other hand, in general [e.g., Figures \ref{images-z}(c) and (f)], the final refocused image does not necessarily maintain the improved DOF of its underlying coherent images. This is due to the detrimental effect diffraction has on the reliability of the retrieved angular information: The one-to-one correspondence between points of the angular sensor $S_b$ and points on the source plane may be compromised by diffraction at the object. Notice that for large enough objects [e.g., Figures \ref{images-z}(g) and (i)], diffraction enters into play only at object distances $z_b>z_a$ that are equal or larger than the DOF of the coherent image; hence, the refocused image has essentially the same DOF of its underlying coherent images [e.g., Figures \ref{images-z}(i)]. 

However, due to the coherent nature of CPI, a complete analysis of the DOF requires accounting for interference. In Figure~\ref{double-slit}, the DOF is obtained by the visibility of the double-slit image: The Rayleigh resolution criterion applied to the image of a double-slit having slit width $a$ and center-to-center distance $d=2a$, gives a visibility $V=10\%$. Based on Figure~\ref{double-slit}, interference between light passing through the two slits limits the DOF both for the coherent and for the final CPI refocused image.

\end{document}